\documentclass{emulateapj}

\lefthead{Bekki}
\righthead{The Galactic globular clusters}

\begin{document}
\title{The Galactic globular cluster system as a fossil record of reionization}

\author{Kenji Bekki\altaffilmark{1}}

\altaffiltext{1}{School of Physics, University of New South Wales,
Sydney 2052, Australia, bekki@bat.phys.unsw.edu.au}

\begin{abstract}

We propose  that structural, kinematical, and chemical properties of
the Galactic globular clusters (GCs) can contain fossil information of 
the cosmic reionization history.
We first summarize possible observational evidences 
for the influence of reionization on the Galactic GC formation.
We then show how structural properties of the GC system (GCS)
in the Galaxy can be influenced by suppression of GC formation
due to reionization
during the Galaxy formation through
hierarchical merging of subgalactic clumps,  
by using numerical simulations with and without suppression of
GC formation by reionization. 
In particular,  we show that if GC formation
in dwarf galaxies that are building blocks of
the Galaxy and  virialized after reionization era ($z_{reion}$)
are completely suppressed,
the present-day radial  distribution of the Galactic GCs depends strongly
on $z_{reion}$.
Our numerical results imply that if GC formation after $z_{reion}$ $\sim$ 15
is strongly suppressed, the origin of the observed structural properties
of the Galactic  GCS can be more naturally explained
in the framework of the hierarchical clustering scenario.

\end{abstract}


\section{Introduction}

The globular cluster system (GCS) of the Galaxy  has long been considered
to contain ``fossil records'' on early dynamical and chemical 
histories of the Galaxy (e.g., Searle \& Zinn 1978;
Harris 1991; Mackey \& Gilmore 2004). 
Therefore physical properties of the GCs, 
such as the abundance gradient (e.g., Searle \& Zinn 1978),
radial density profile (e.g., Zinn 1985; van den Bergh 2000), 
kinematics (Freeman 1993), and proper motion (Dinescu et al. 1999)
have been extensively discussed in the context of the Galactic
formation history.
Although recent numerical simulations of the Galaxy formation
based on the cold dark matter (CDM) model have provided
a model explaining 
both dynamical and chemical properties of 
the Galactic old stellar halo (Bekki \& Chiba 2000, 2001),
theoretical models trying to explain the 
observed structural, kinematical, and chemical
properties of the Galactic GCS {\it in a self-consistent manner}  
have not yet been provided.
This is partly because 
physical conditions required for globular cluster (GC) formation
in low mass dwarfs, which can be the building blocks
of the Galaxy in the hierarchical clustering scenario,
have not yet been completely clarified.

Recently, effects of reionization 
on galaxy formation have been extensively discussed,
in particular, 
in the context  of star formation histories and
old stellar populations in low mass dwarfs 
(e.g.,  Bullock et al 2000; Gnedin 2000;
Susa \& Umemura 2004; 
Grebel \& Gallagher 2004; Bekki \& Chiba 2005). 
These theoretical works suggested that 
suppression of star formation by heating and  gas loss
resulting from reionization can be a very important
physical process for better understanding the observability
and the early star formation activities in dwarfs,
though Grebel \& Gallagher (2004) did not find
any clear fossil evidences  that support
the occurrence of such suppression at reionization era ($\sim$ 12.8 Gyr) 
in nearby dSphs. 
Although it has already been discussed whether 
reionization can trigger (Cen 2001)
or suppress (Santos 2003) the formation of globular clusters,
it remains totally unclear how reionization determines 
dynamical and chemical properties of the Galactic GCS.
Given the fact that unprecedentedly wealthy data sets 
on physical properties of the Galactic GCs are now
available (e.g., Harris 1996; Mackey \& Gilmore 2004), 
it is worthwhile to discuss  possible effects of reionization
on the GC formation by comparing the observations
with theoretical results in order to better understand the origin
of the Galactic GCS.

The purpose of this Letter is to  demonstrate, for the first time,
that physical properties
of the Galactic GCS can be significantly influenced by reionization
by comparing observational data of the Galactic GCS with numerical
simulations including possible influences of reionization on
GC formation.
We particularly show how the radial  profile of
the number distribution of the Galactic GCS can
be influenced by {\it the suppression of GC formation
by reionization } by using numerical simulations of the Galaxy formation 
based on the cold dark matter (CDM) model.
We firstly summarize three key observational results which
can be interpreted as possible evidences for GC formation 
influenced by reionization in \S 2, 
and then describe our  numerical results in \S 3 and \S 4.
We suggest several observable GC  properties that can be used
to discuss reionization effects on GC formation for disk galaxies
in \S 5.

\section{Three key observational results}

Interpretations given below for observational results 
would be somewhat speculative and
are not unique,
because there can be 
several alternative interpretations for each of the results.
We however suggest that the origin of the results
can be closely associated with some  physical processes 
(reionization etc.)
influencing the early Galactic GC formation.
It should be also stressed that the following
observations  could  be only three among many  of   
fossil evidences for the impact of reionization
on the Galactic GC formation.

First is the steeper slope in the number density profile
of the Galactic GCS for the outer Galactic halo (van den Bergh 2000).
As Figure 1 reveals, showing the radial profile
of the cumulative number distribution of
the GCS ($N_{\rm GC}(R)$),  the GC distribution 
(${\rho}_{\rm GC}(R)$) becomes progressively
steeper toward the outer part of the Galaxy. If ${\rho}_{\rm GC}(R)$
is approximated as 
${\rho}_{\rm GC}(R) \propto R^{n}$, 
where $R$ is the distance of a GC from
the Galactic center, $n$ is roughly $-2.2$ for
2 $<$ $R$(kpc) $<$ 8 kpc (van den Bergh 2000) 
and $-3.0$ for 20 $<$ $R$(kpc) $<$ 100 kpc 
(It should be noted here that if $N_{\rm GC}(R)$ is  
approximated as $N_{\rm GC}(R) \propto R^{\alpha}$, 
then $\alpha \approx 3+n$).
The half-number radius of the Galactic GCS ($R_{\rm h, GC}$) can be estimated
as 5.2 kpc.
Recent high-resolution cosmological simulations 
have shown that the cumulative radial distribution of dark matter subhalos
in a Milky Way sized galaxy can increase linearly
with $R/R_{200}$, where $R_{200}$ corresponds to the virial radius
of the halo,
and have a half-number radius of $\sim$ $0.6 \times R_{200}$
(See Figure 2 in Willman et al. 2004).

These simulation results 
mean that the radial number distribution ${\rho}_{\rm shalo}(R)$
of subhalos can be approximated as 
${\rho}_{\rm shalo}(R) \propto R^{s}$ with $s$ $\approx$  $-2$ 
and have the half-number radius ($R_{\rm h, shalo}$) of 209 kpc
for $R_{200}$ = 348 kpc that is a reasonable value for
the Galactic  mass of $\approx$ $2.0$ $\times$ $10^{12}$ $M_{\odot}$
(Wilkinson \& Evans 1999) in the NFW profile
(Model 5 in Navarro et al. 1996). 
These observational and theoretical results raise the following
question: 
If most of the Galactic GCs originate from satellite
galaxies in the subhalos 
(e.g., through accretion and tidal stripping of GCs),
how can we explain  the above  remarkable differences
{both in the slope of density profiles
in the Galactic outer region 
($n$ and $s$) 
and in the half-number radius
($R_{\rm h, GC}$ and $R_{\rm h, shalo}$)}
between the GC distribution  and the simulated subhalo one ? 
A possible explanation for this is that GC formation in 
outer subhalos (i.e., low mass dwarfs) are strongly suppressed 
by some physical mechanisms.

Second is the apparently clear bimodality in the metallicity distribution
of the GCS (e.g., Zinn 1985; Reed et al. 1994).
Theoretical studies suggested that
the apparently clear secondary peak in the [Fe/H] distribution
can be due to secondary dissipative merger events that can
form new, metal-rich GCs (MRCs) in the early dynamical history
of the Galaxy (e.g., Ashaman \& Zepf 1992; 
Bekki \& Chiba 2002; Bekki et al. 2002).
However recent semi-analytic study based on the CDM model
(Beasley et al. 2002)
has demonstrated that  
such clear gaps between the two peaks seen in galaxies
can not be easily explained
by such merger-induced GC formation {\it without truncating
the formation of metal-poor GCs at very high redshifts ($z$ $\sim$ 5)}. 
Therefore the observed bimodality 
can possibly imply that the strong suppression of the formation 
of the Galactic metal-poor GCs (MPCs) by some physical processes.

Third is the significant difference in GC number between the Galaxy
and M31 (e.g., van den Bergh 2000).
Although the total mass can be roughly the same between
the Galaxy and M31 (Evans et al. 2000),
the total number of GCs in the Galaxy  (160 $\pm$ 20)
is much smaller than that in M31 (400 $\pm$ 55; van der Bergh 2000).
This significant difference in the formation efficiency of GCs  per galaxy mass
can possibly indicate that GC formation is {\it globally} 
suppressed 
in the Galaxy compared with M31 owing to some physical processes.
Thus we can interpret the above three observations as possibly implying 
the suppression of GC formation in the early Galactic history. 
In the followings, we adopt a working hypothesis that
reionization can be the mechanism responsible for the suppression
of GC formation,
and thereby investigate how it influences structural  properties
of the Galactic GCS.

\section{The numerical model}

We simulate the formation of a Milky Way sized galaxy halo
in a $\Lambda$CDM Universe 
with $\Omega =0.3$, 
$\Lambda=0.7$, $H_{0}=70$ km $\rm s^{-1}$ ${\rm Mpc}^{-1}$,
and ${\sigma}_{8}=0.9$,
and thereby investigate merging/accretion
histories of subhalos that can contain low mass dwarfs with GCs.
The way to set up initial conditions for the numerical simulations
is essentially
the same as that adopted by Katz \& Gunn (1991) and Steinmetz \& M\"uller
(1995). 
We consider an isolated homogeneous, rigidly rotating sphere, on which
small-scale fluctuations according to a CDM power spectrum are superimposed.
The initial total mass, radius, spin parameter ($\lambda$),
and the initial overdensity ${\delta}_{i}$   of the sphere
in the present model are $6.0\times10^{11}\rm M_{\odot}$, 30 kpc, 
0.08, and 0.26, respectively. 
Initial conditions similar to those adopted in the present study
are demonstrated to be plausible and realistic for the formation
of the Galaxy (e.g., Steinmetz \& M\"uller 1995; Bekki \& Chiba 2001).

We start the collisionless simulation at $z_{\rm start}$ (=30) and follow it 
till $z_{\rm end}$ (=1) to identify virialized subhalos
with the densities larger than $170 {\rho}_{\rm c}(z)$,
where ${\rho}_{\rm c}(z)$ is the critical density, 
at a given $z$. 
The minimum number of particles within a virialized halo
($N_{\rm min}$) is set to be 32 corresponding to the mass resolution
of 3.8 $\times$ $10^{7}$ $M_{\odot}$. 
For each individual virialized subhalo
with the virialized redshift of $z_{\rm vir}$,
we estimate a radius ($r_{\rm b}$) within which 20 \%  of the total mass
is included,  and then the particles within $r_{\rm b}$ are labeled 
as ``baryonic''  particles.
20 \% of the  {\it outermost}  baryonic particles 
in a subhalo are  labeled as ``GC particles''
so that the total number of GCs in the halo is  1 
for the smallest halo with $N_{\rm min}$ =32.
Thus the present simulation {\it assumes} that GCs can be formed
in all low mass subhalos.

In order to investigate the effects of the suppression
of GC formation via reionization on the final structural
properties of the simulated GCS,
we adopt the following somewhat idealized 
assumption: {\it If a subhalo is virialized after the completion
of the reionization ($z_{\rm reion}$),
GC formation is totally suppressed in the halo.}
Therefore, ``GC particles'' in
the subhalo with $z_{\rm vir}$ $<$ $z_{\rm reion}$  
are not considered at all in the estimation of structural properties
of the simulated GCS.
Recent WMAP ({\it Wilkinson Microwave Anisotropy Probe})
observations have shown that plausible $z_{\rm reion}$  
ranges from 11 to 30 (Spergel et al. 2003; Kogut et al. 2003)
whereas quasar absorption line studies can give the lower limit
of 6.4 for $z_{\rm reion}$ (Fan et al. 2003).
Guided by these observations, we investigate the models
with $z_{\rm reion}$  = 0 (no reionization), 5, 10, and 15.

Thus the present collisionless simulations follows
the dynamical evolution of GCs originating from subhalos 
and do not include
any {\it dissipative} formation of GCs associated with
starbursts during merging/accretion
of subgalactic halos.
Therefore, the simulated GCs can mimic old,  MPCs
rather than MRCs (e.g., ``disk GCs'' in the Galaxy).
All the calculations of the Galaxy  formation
have been carried out on the GRAPE board (Sugimoto et al. 1990).
Total number of particles used in a simulation is 508686 
and the gravitational softening length is 0.38 kpc.
We consider that the final structure of the simulated
GCS at $z_{\rm end}$ (=1) is the same as that at $z$ = 0,
because the GCS is dynamically relaxed completely  until $z_{\rm end}$ 
for the present isolated model in which recent accretion/merging
of dwarfs do not happen after $z_{\rm end}$. 
We used the COSMICS (Cosmological Initial Conditions and
Microwave Anisotropy Codes), which is a package
of fortran programs for generating Gaussian random initial
conditions for nonlinear structure formation simulations
(Bertschinger 1995).

\section{Result}

Figure 2 shows how the Galactic GCS can be formed through
merging of subgalactic halos with GCs in the model with $z_{\rm reion}$
= 10.
Subhalos grown from higher peaks of CDM density fluctuation
are virialized earlier ($z_{\rm vir}$ $>$ $z_{\rm reion}$)  
to contain older GCs within them,
and these subhalos then merge with one another to form 
bigger subhalos with a larger number of GCs ($z$ = 5.2 in Figure 2). 
Numerous small subhalos without GCs can be seen around this redshift,
which reflects the fact that
about 47 \% of all subhalos 
are formed  later than $z_{\rm reion}$ from lower density peaks
and thus suffer suppression of GC formation.
These subhalos with and without GCs 
continue to merge ($z$ = 3.0) to finally form
a single massive halo ($z$ = 2.1) with a number of bigger
satellite halos with GCs.
After accretion/merging of satellite subhalos
and the resultant dynamical relaxation,
a GCS finally forms  with a relatively smooth GC density distribution,
a triaxial shape, and the major axis of the distribution aligned with
that of the dark matter halo. 

Figure 3 shows the final radial profile of the cumulative
number distribution ($N_{\rm GC}(R)$) of the simulated GCS in the model with
$z_{\rm reion}$ = 0, 10, and 15.
It is clear form this figure that the profiles in the models
with reionization  are closer to
the observed one than that in the model 
without reionization ($z_{\rm reion}$ = 0).
In particular, the profile in the  model 
with higher $z_{\rm reion}$ (=15)
can more closely resemble the observed one 
than that in the model with lower $z_{\rm reion}$ (=10). 
These results imply that the formation of 
the radial distribution of the Galactic
GCS could have been influenced by reionization.
Later accretion/merging  of younger subhalos with GCs
can add new GCs to the outer halo of the simulated Galaxy
and thus flatten the density profile of the GCS. 
In the model with $z_{\rm reion}$ = 15, 
accretion/merging of younger subhalos can not
influence the distribution of the GCS, 
because they do not contain GCs owing to later 
virialization with $z_{\rm vir}$ $<$ $z_{\rm reion}$
(i.e., suppression of GC formation).
This is the essential reason why the model with $z_{\rm reion}$ = 15
can better reproduce the observation.
The observed small half-number radius ($R_{\rm h, GC}$) 
of 5.2 kpc (e.g., van den Bergh 2000)
can be also better reproduced by the model with $z_{\rm reion}$ = 15
showing  $R_{\rm h, GC}$ of 5.0 kpc.

The global mass profile of the simulated Galactic dark matter halo
does not depend on $z_{\rm reion}$ in the present study.
Therefore, Figure 3 also suggests that
the half-number radius of a GCS ($R_{\rm h, GC}$) in a model 
with respect to the half-mass radius ($R_{\rm h, dm}$)
of the dark matter halo is smaller for higher $z_{\rm reion}$. 
Given the fact that ongoing observations on kinematics of
GCSs in galaxies enable us 
to infer mass profiles of the dark matter halos (e.g., Bridges et al. 2003),
this $R_{\rm h, GC}$-$R_{\rm h, dm}$ ratio can be regarded as 
an important observable property that measures the influence of reionization
on GC formation in galaxies.
As expected from the adopted assumptions of the model,
the total number of GCs ($n_{\rm GC}$)  depends strongly on $z_{\rm reion}$
in such a way that $n_{\rm GC}$ is larger for smaller $z_{\rm reion}$ 
(e.g., $n_{\rm GC}$ = 4986 and 116 for $z_{\rm reion}$ 10 and 15, respectively).
This result might well provide a clue to a question
as to why the Galaxy has only $<$  200 GCs.
Thus the observed compact GC distribution with a radius-dependent
slope of the distribution in the Galaxy can be better explained
by the model with $z_{\rm reion}$ = 15. 
 
\begin{figure}
\plotone{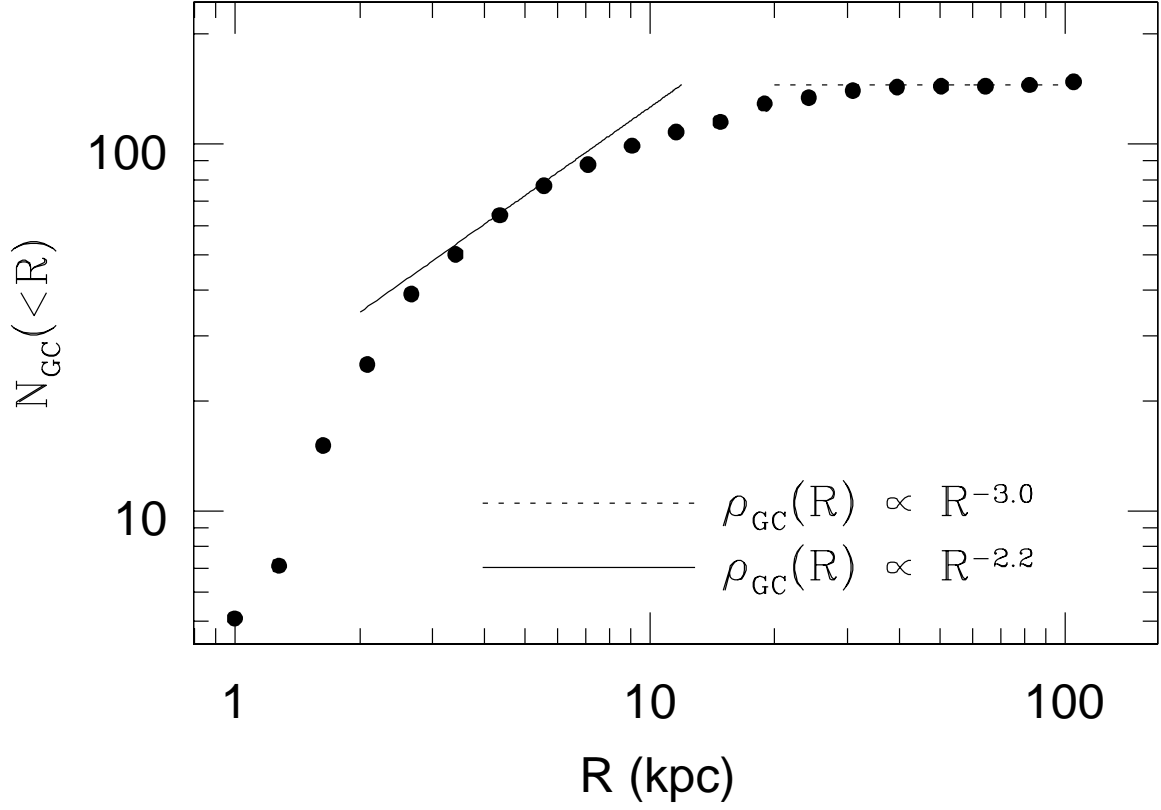}
\caption{
The observed cumulative distribution ($N_{\rm GC}(R)$) 
of the Galactic GCs, where $R$ is the distance from the
Galactic center. Data sets for 150 GCs in Harris (1996)
are used here. The solid line represents
the best-fit GC number density profile of  
${\rho}_{\rm GC}(R)$ $\propto$ $R^{-2.2}$
for 2 $<$ $R$(kpc) $<$ 8 (van den Bergh 2000).
In order to show the difference in the slopes of
the GC number density profile between
the inner and the outer part of the Galaxy,
the profile of ${\rho}_{\rm GC}(R)$ $\propto$ $R^{-3.0}$ 
is also shown by a dotted line for  20 $<$ $R$(kpc) $<$ 100.
\label{fig-1}}
\end{figure}

\placefigure{fig-1}
\placefigure{fig-2}

\placefigure{fig-3}
\placefigure{fig-4}

\section{Discussions and conclusions}

Previous theoretical studies demonstrated  that
destruction of  GCs is a 
key physical process that controls the radial profiles
of GCSs in galaxies 
(e.g., Baumgardt 1998; Fall \& Zhang 2001; Vesperini et al. 2003).
In order to discuss how  the simulated radial profiles of GCSs
can be modified by this destruction process, we
numerically investigated the orbital evolution of GCs in the model with
$z_{\rm reion}=15$ and thereby estimated the orbital eccentricity
($e$) and pericenter ($r_{\rm p}$) of each of the GCs formed before
reionization. We found that most of the GCs have highly radial orbits
($e \sim 0.6-0.8$) with smaller pericenter distances ($r_{\rm p} \le$ 10kpc).
Aguilar et al. (1988) demonstrated that GCs with highly eccentric
orbits ($e > 0.8$) and small pericenter distances ($r_{\rm p} \le$ 2kpc) 
can be completely destroyed by the Galactic tidal field well within
the Hubble time. By combining this previous result with the above
one on orbital properties of the simulated GCs, we found
that about 55 \% of the inner GCs in the simulation 
can be destroyed by the Galactic tidal field.

Given the fact that selective destruction of GCs with higher $e$
and smaller $r_{\rm p}$ 
can result in the flattening of the radial profiles of GCSs
(e.g., Vesperini et al. 2003),
the above result implies that the radial density profile of the
{\it initial} Galactic GCS possibly composed mostly of 
GCs formed before $z_{\rm reion}$ can be significantly steeper than that observed
now. 
Furthermore the kinematics of the Galactic GCS 
is observationally suggested  to be 
fairly isotropic (e.g., Freeman 1993), which appears to be 
inconsistent of the simulated kinematics of the GCS composed
mostly of GCs with highly eccentric orbits.  
This apparent inconsistency implies that the origin of
the observed isotropic kinematics can be also closely associated
with later destruction processes of GCs.

The present study neglects the influence
of the {\it local} UV field from the first-generation GCs
(and very massive stars) formed
in subgalactic clumps at $z > z_{\rm reion}$ in a protogalaxy
on GC formation at $z >z_{\rm reion}$ in the protogalaxy.
The simulated GCS at $z=16$ has a 
half-number radius of $\sim 50$kpc,
and an order of magnitude estimation implies that 
GC formation within the half-number radius might well be 
influenced by local UV effects from population III stars
even before $z_{\rm reion}$.
Accordingly structural and kinematical properties of the inner GCs,
which can be formed well before $z_{\rm reion}$,
could be particularly modified by this local effect.
More quantitative estimation of  this local UV
effects on GC formation can  be done in 
our future numerical studies with more realistic and sophisticated
models of GC formation. These future studies will also address
a role of reionization in shaping the bimodal color distribution
of the Galactic GCS.

\acknowledgments
We are  grateful to the anonymous referee for valuable comments,
which contribute to improve the present paper.

\clearpage


\clearpage

\clearpage


\clearpage
\plotone{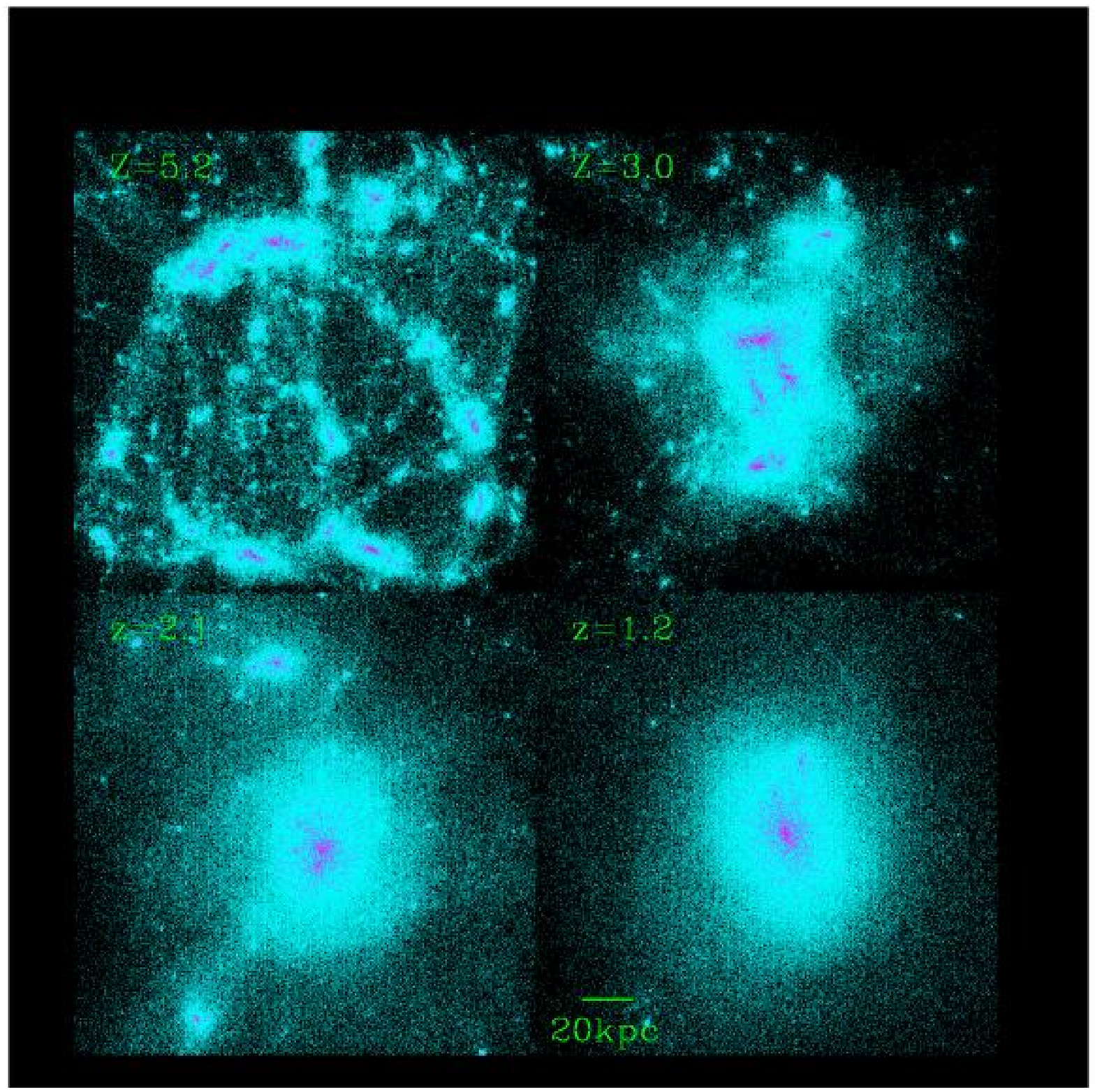}
\clearpage

\begin{figure}
\epsscale{.70}
\vspace{15mm}
\caption{
Time evolution of mass distributions for the dark matter
halo (cyan) and GCs (magenta) in the simulated Galaxy.
Note that the Galactic GCS can be formed from merging
of smaller subgalactic clumps with and without GCs.
\label{fig-2}}
\end{figure}

\clearpage

\begin{figure}
\plotone{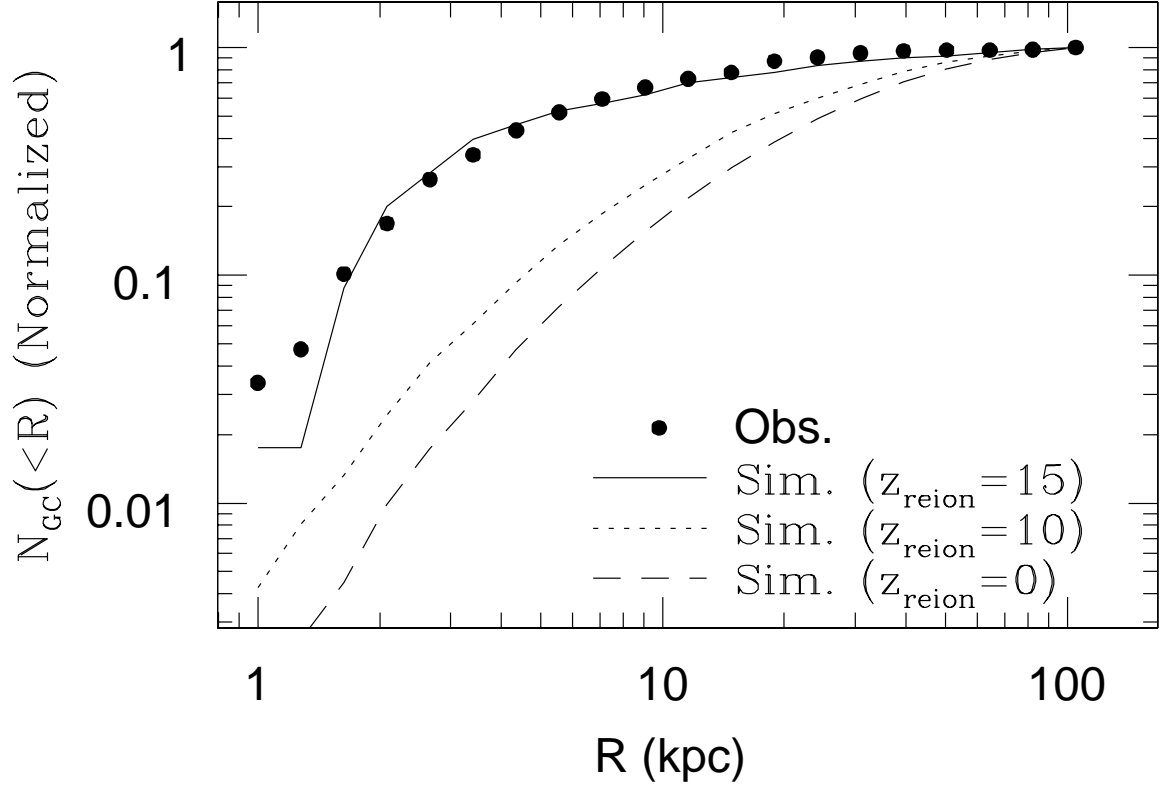}
\caption{
Comparison between the observed (normalized) cumulative number distribution
of GCs (filled circles) and the simulated one corresonding to $z=0$
for the models with
$z_{\rm reion}$ = 0 (dashed),  10 (dotted), and 15 (solid).
$z_{\rm reion}$ = 0 means that no suppression of GC formation by
reionization is assumed.
Note that the simulated profile
in  the model with higher $z_{\rm reion}$ (=15) can
be much closer to the observation than those in the models
with smaller $z_{\rm reion}$ (=0 and 10). 
\label{fig-3}}
\end{figure}

\end{document}